\documentclass[]{aipproc}
\layoutstyle{8x11double}
\def\(({\left(}
\def\)){\right)}
\def\[[{\left[}
\def\]]{\right]}
\begin{document}

\title[Temperature chaos]
      {Rejuvenation and Memory in Migdal-Kadanoff Spin Glasses}
\author{M. Sasaki}{
  address={Institute for Solid State Physics, University of Tokyo, 
           Kashiwa-no-ha 5-1-5, Kashiwa, 277-8581, Japan.}
}

\author{O.C. Martin}{
  address={Laboratoire de Physique Th\'eorique et Mod\`eles Statistiques,
           b\^at. 100, Universit\'e Paris-Sud, F--91405 Orsay, France.}
}

\begin{abstract}
We study aging phenomena of Migdal-Kadanoff spin glasses 
in order to clarify relevancy of temperature chaos to 
rejuvenation and memory. By exploiting renormalization, 
we do efficient dynamical simulations in very wide 
time/length scales including the chaos length. 
As a consequence, we find that temperature chaos 
and temperature dependence of speed of equilibration 
cause two significantly different effects against 
temperature variations, i.e., rejuvenation for {\it positive} temperature 
variation and memory for {\it negative} temperature variation, 
as are observed experimentally in spin glasses. 
\end{abstract}

\date{\today}

\maketitle
It is well known that in spin glasses, dynamical effects strongly depend on 
the history of the system after quench from above the transition 
temperature $T_{\rm c}$. These phenomena are called aging and 
have been studied using various experimental 
protocols~\cite{VincentHammann00,NordbladSvedlindh98,JonssonYoshino03}. 
Measurement of ac-susceptibility 
during $T$-cycle~\cite{VincentBouchaud95}, 
which is employed in this work, is one of them. 
This experiment consists of the following three stages. 
In the first stage, the system 
is quenched from above $T_{\rm c}$ and it is kept at 
a temperature $T$ ($<T_{\rm c}$) during a time $t_1$. 
In the second stage the temperature is temporarily changed 
to $T\pm \Delta T$ ($<T_{\rm c}$) during a time $t_2$, 
and then it is set back to $T$ in the third stage. 
Ac-susceptibility $\chi$ is measured during all the three stages. 
In the case $-\Delta T$, $\chi$ in the third stage 
resumes its relaxation from the value at the end of the period $t_1$ 
as if the system remembers how far the relaxation 
at $T$ had proceeded before the perturbation 
({\it memory effect}). On the other hand, the system is {\it rejuvenated} by 
positive $T$-cycle, and we observe strong relaxation of $\chi$ 
in the third stage. 

From a theoretical point of view, ``temperature chaos''~\cite{BrayMoore87}, 
decorrelation of the equilibrium states at two temperatures 
beyond the so-called chaos length $\ell(T,T')$, 
has been one of the most conceivable causes of rejuvenation. 
However, temperature chaos seems to be incompatible with memory 
if one naively thinks. Therefore, relation among 
temperature chaos, rejuvenation and memory has been 
vigorously studied~\cite{YoshinoLemaitre00,JonssonYoshino03} 
as a key to understand aging phenomena. 
In this work, we also address this issue by 
studying Migdal-Kadanoff (MK) spin glasses. 
There are mainly two merits in working on MK spin glasses. 
First the existence of temperature chaos is shown in this model, 
and whose exact renormalization~\cite{SouthernYoung77} allows one 
to measure the chaos length $\ell(T,T')$. Second we can do 
efficient dynamical simulations at very long {\it time} scales
by exploiting renormalization~\cite{SasakiMartin03}. 
(See also~\cite{Scheffler03} as a similar approach.) 
In the previous work~\cite{SasakiMartin03}, we have 
investigated dynamics of MK spin glasses 
by utilizing these advantages, and found that temperature chaos 
causes rejuvenation, but it also destroys most of the memory 
if the length scale equilibrated during the second stage 
is much larger than $\ell(T,T')$. 
The main purpose of this work is to show that memory is preserved 
if we take temperature dependence of speed of equilibration 
into account. 

The outline of the paper is as follows. First, we introduce MK spin glasses. 
Second, we briefly recall temperature chaos in MK spin glasses. 
Third, we explain how renormalized dynamics can be used to probe
rejuvenation and memory on very wide time/length scales. 
Finally, we show results of the both positive and negative 
$T$-cycle simulations. The last section is 
devoted for discussion and conclusions. 

\begin{figure}[t]
\includegraphics[angle=0,width=\columnwidth]{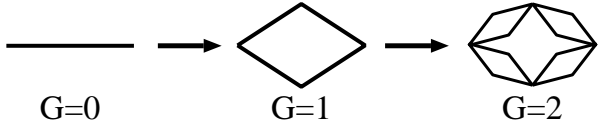} 
\caption{Construction of a MK lattice ($b=2$).}
\label{fig:mk}
\end{figure}

\paragraph*{The model ---}
We consider MK lattices following 
the standard real space renormalization group
approximation~\cite{SouthernYoung77} to the
Edwards Anderson (EA) model~\cite{EdwardsAnderson75}.
The recursive construction of such hierarchical lattices is 
described in Fig.~\ref{fig:mk}; edges are 
replaced by $2 b$ edges so the ``length'' of the lattice
is multiplied by $2$. We call generation ``level'' the order
of the recursion and $G$ 
the total number of these. Then the lattice length $L$ is
$2^G$ and the number of bonds is $(2 b)^G$ (which is also
roughly the number of sites); one can thus identify
$1 + \ln b / \ln 2$ with 
the dimension of space on such a lattice.
When all the edges are constructed, each is assigned
a random coupling $J_{ij}^0$. The superscript ``0'' 
implies that $J_{ij}^0$ are bare couplings. 
Similarly, on each site $i$ we put an Ising spin $S_i = \pm 1$. 
The Hamiltonian is
\begin{equation}
\label{eq:H}
H_J(\{S_i\}) = - \sum_{<ij>} J_{ij}^0 S_i S_j,
\end{equation}
where the sum is over all the nearest 
neighbor spins of the lattice. 
%
All of the work presented here will be for 
three dimensions ($b=4$) with couplings $J_{ij}^0$ taken
from a Gaussian of mean $0$ and width $1$.
The model then undergoes a spin glass transition
at $T_c \approx  0.896$~\cite{NifleHilhorst92}.

\paragraph*{Temperature chaos in MK spin glasses ---}
A great advantage of MK lattices is that 
it allows us to do renormalization procedure exactly~\cite{SouthernYoung77}. 
Now let us denote a set of spins of level $n$ as $\{ S_n \}$. 
By tracing over spins of lower levels, we explicitly obtain 
\begin{eqnarray}
\hspace*{-4mm}P_k(\{S_0\},\cdots,\{S_{G-k}\})
\hspace{-3.5mm}&\equiv&\hspace{-3.5mm}
\frac{{\rm Tr}_{\{S_{G-k+1} \}\cdots\{S_{G} \}} \exp(-{\cal H}_{J}^0)}
{{\rm Tr}_{\{S_{0} \}\cdots\{S_{G} \}} \exp(-{\cal H}_{J}^0)} \nonumber \\
\hspace{-3.5mm}&\propto& \hspace{-3.5mm}\exp[-{\cal H}_{J}^{k}(\{S_0\},\cdots,\{S_{G-k}\})],
\end{eqnarray}
\begin{equation}
{\cal H}_{J}^{k}(\{S_0\},\cdots,\{S_{G-k}\})=- \sum\nolimits_{<ij>_{G-k}} 
{\tilde J}_{ij}^{k} S_i S_j,
\end{equation}
where the sum $\sum_{<ij>_{n}}$ is over all the nearest 
neighbor spins of a MK lattice with generation $n$.
Effective couplings ${\tilde J}_{ij}^{k}$ are 
$J_{ij}^{0}/T$ for $k=0$. Otherwise, they are calculated by 
the recursion formula
\begin{equation}
{\tilde J}_{ij}^{k+1}=\sum_{l=1}^4 {\rm arctanh}[
\tanh {\tilde J}_{il}^{k}\tanh {\tilde J}_{lj}^{k}].
\end{equation}
In this equation, ${\tilde J}_{il}^{k}$ and 
${\tilde J}_{lj}^{k}$ lie on the $l$-th path 
connecting $i$ and $j$. 

Now let us turn to temperature chaos in MK spin glasses. 
If we start from the same set of bare couplings, 
$\{ {\tilde J}_{ij}^{0}(T) \}$ and 
$\{ {\tilde J}_{ij}^{0}(T+\Delta T) \}$ are completely correlated 
for any $T$ and $\Delta T$. Does the correlation still survive after 
renormalization is repeated again and again? This question was 
first addressed by Banavar and Bray~\cite{BanavarBray87}, and 
they have found that for arbitrarily small $\Delta T$ 
$\{ {\tilde J}_{ij}^{k}(T) \}$ and 
$\{ {\tilde J}_{ij}^{k}(T+\Delta T) \}$ 
become completely decorrelated when $k$ is large enough, 
indicating that spin polarizations at two temperatures are 
different at all. In table~\ref{tab:Cvalues}, we show 
the linear correlation coefficient 
\begin{equation}
C(L,T,T') = \frac{\overline{ {\tilde J}^{k}(T) 
{\tilde J}^{k}(T')}}
{\sigma(T) ~~ \sigma(T')},
\label{eq:correlation_coeff}
\end{equation}
for $T=0.7$ and $T=0.65$. In this definition, $L=2^k$, 
${\overline{ \cdots }}$ is the disorder average,
$\sigma$ is the standard deviation of ${\tilde J}^{k}$,
and we have used the fact that $\overline{ {\tilde J}^{k}}=0$. 
We see $C(L,T,T')$ rapidly drops from $1$ to $0$ around 
$L\approx 2^{11}$. We hereafter utilize $C(L,T,T')$ as an 
indicator of temperature chaos, 
and define the chaos length $\ell(T,T')$ as the value of $L$ where 
$C=1/{\rm e}$.  (In this case, $\ell(T,T')\approx 2^{12}$.)
\begin{table}
\begin{tabular}{ccccccc} 
\hline
  \tablehead{1}{c}{b}{$L$}
 &\tablehead{1}{c}{b}{$2^5$}
 &\tablehead{1}{c}{b}{$2^7$}
 &\tablehead{1}{c}{b}{$2^9$}
 &\tablehead{1}{c}{b}{$2^{11}$}
 &\tablehead{1}{c}{b}{$2^{13}$} 
 &\tablehead{1}{c}{b}{$2^{15}$} \\ \hline
C & 1.00 & 0.99 & 0.92 & 0.62 & 0.10 & 0.00 \\
\hline
\end{tabular}
\caption{Size dependence of $C(L,T=0.7,T'=0.65)$.}
\label{tab:Cvalues}
\end{table}
\paragraph*{Exploiting renormalization for dynamics ---}
Since our purpose is to examine relevancy of temperature chaos on dynamics, 
we should do dynamical simulations at very long time so that 
the length scale equilibrated during simulation is 
comparable with the chaos length. However, this condition is hardly 
satisfied by usual Monte-Carlo simulations because 
frustration and randomness inherent in spin glasses 
make their dynamics extremely slow. 
In fact, measurements of 
$L_{\rm eq}(t)$, the equilibrated length during $t$, 
in 3d-EA spin glass model have shown that $L_{\rm eq}(t)$ 
for $t= 10^{6}$ Monte Carlo Sweeps (MCS) is less than $10$ 
at any temperatures below $T_{\rm c}$~\cite{KomoriYoshino99}, 
while it is almost impossible to go beyond the time scale 
by the present computer resource. 
This length scale seems to be hopelessly shorter than the 
chaos length. (Recall that $\ell(T,T')\approx 2^{12}$ in the previous case 
though the temperature difference is not so small.) 

In order to overcome the difficulty, we exploit renormalization 
for dynamics~\cite{SasakiMartin03}. The basic idea is as follows. 
Suppose we focus on a time window 
$t_{\rm min}\le t \le t_{\rm max}$. Between $t=0$ and
$t=t_{\rm min}$ the system has had time to equilibrate up to
the length scale $l(t_{\rm min})$; essentially all out of equilibrium
physics comes from larger length scales. On MK lattices, this means
that the spins whose generation ``level'' is larger than $G_{\rm min}$ 
(with $2^{\rm NRG} = l(t_{\rm min})$ and ${\rm NRG}\equiv G-G_{\rm min}$) 
are in local equilibrium; the other spins have dynamics that is 
well described by the effective Hamiltonian 
${\cal H}_{J}^{{\rm NRG}}(\{S_0\},\cdots,\{S_{G_{\rm min}}\})$. 
In practice, we adopt the following procedure to implement 
this idea with taking temperature dependence of 
speed of equilibration into account. 
\begin{itemize}
\item[1.] Calculate the effective couplings at $T_{\rm H}$ 
(the higher temperature used in T-cycle protocol) and 
those at $T_{\rm L}$ (the lower one). We first generate 
a large number of bare couplings from a Gaussian 
of mean $0$ and width $1$. Then, we do renormalization procedure 
to produce an ensemble of effective couplings. This process is iterated 
$\textrm{NRG}$ times. The final effective couplings are then 
randomly assigned to the edges of a MK lattice 
of size $2^{G_{\rm min}}$. 
\item[2.] The direction of each spin at $t=0$ is chosen randomly 
with equal probability, corresponding to
a quench from an infinitely high temperature at an infinite rate. 
\item[3.] At $T_{\rm L}$, 
we simply do standard Monte Carlo by using 
${\cal H}_{J}^{{\rm NRG}}(\{S_0\},\cdots,\{S_{G_{\rm min}}\})$ 
prepared at step 1. 
\item[4.] At $T_{\rm H}$, dynamics is further accelerated by the 
following procedure. We first calculate 
${\cal H}_{J}^{{\rm NRG}'}(\{S_0\},\cdots,\{S_{G'_{\rm min}}\})$, 
where ${\rm NRG}'= G-G_{\rm min}'$ and ${\rm NRG}' > {\rm NRG}$. 
Then we do Monte Carlo by using ${\cal H}_{J}^{{\rm NRG}'}$ to update 
the spins whose level is smaller (or equal to) $G_{\rm min}'$. 
After each MCS, the lower spins $\{ S_{k} \}$ 
$(G_{\rm min}'<k\le G_{\rm min})$ are 
locally equilibrated with fixing the spins of smaller levels. 
\end{itemize}
Note that one MCS at $T_{\rm L}$ ($T_{\rm H}$) on the renormalized
lattice corresponds to a huge number of sweeps
on the non-renormalized lattice, in fact to the
number needed to equilibrate on the length scale $2^{\textrm{NRG}}$ 
($2^{\textrm{NRG}'}$). 


\paragraph*{Results ---}
We use the standard temperature cycling protocol and measure
a quantity similar to the ac-susceptibility defined as~\cite{KomoriYoshino00}
\begin{equation}
\chi(\omega,t)=\frac{1-Q(t+\frac{2\pi}{\omega},t)}{T},
\end{equation}
where $Q(t,t')\equiv \sum_i \langle W_i S_i(t) S_i(t') \rangle / (\sum_i W_i)$ 
and $W_i$ is the weight of site $i$. Our results concentrate 
on the choice that $W_i$ is proportional to the connectivity of site $i$. 
However, our additional simulations of the case $W_i=1$ have shown that 
the both choices lead to the same results qualitatively. 
Every MCS updates all the spins once.
An alternative choice is to sweep the bonds, updating their end spins
as in~\cite{RicciRitort00}; we 
have checked that the results are hardly affected by the method used. 
In the both positive and negative $T$-cycle simulations, 
we use $T_{\rm H}=0.7$ and $T_{\rm L}=0.65$. 
Note that these two temperatures are the same as those 
in Table~\ref{tab:Cvalues}. 
The period ${2\pi}/{\omega}$ of ac-field is 16 MCS. 
All the simulations are done on  MK lattices with 
five generations ($G_{\rm min}=5$) using $0\le \textrm{NRG} \le 15$. 
Since we calculate renormalized couplings 
at $T_{\rm L}$ and $T_{\rm H}$ from the {\it same} set of bare couplings, 
they are highly correlated when $\textrm{NRG}$ is small. However, 
their correlation vanishes for large $\textrm{NRG}$ due to temperature chaos. 
The difference ${\rm NRG}'-{\rm NRG}$ is $1$ for all the simulations. 
We hereafter denote $\chi$ with a negative (positive) $T$-cycle 
as $\chi_{\rm Ncycle}$ ($\chi_{\rm Pcycle}$) 
and the isothermal $\chi$ at $T$ as $\chi_{\rm iso}(T)$. 
\begin{figure}[ht]
\includegraphics[angle=0,width=\columnwidth]{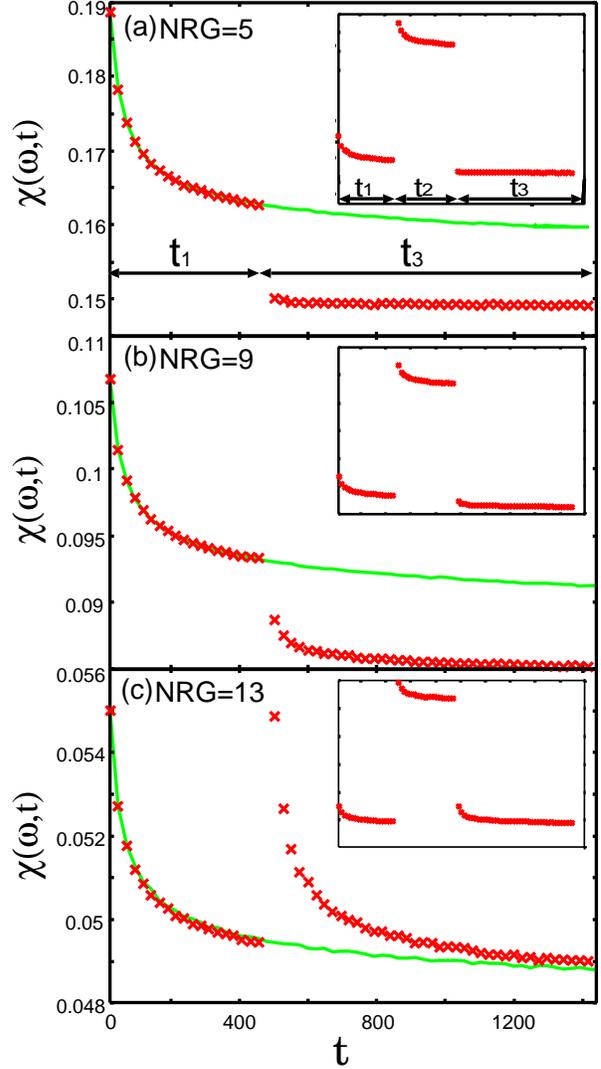} 
\caption{$\chi(\omega,t)$ with a positive $T$-cycle (crosses) 
for $\textrm{NRG}=5$, 9 and 13 (inset). 
Temperature is temporarily increased from $T_{\rm L}=0.65$ to 
$T_{\rm H}=0.7$ during the period $t_2$. In the main frame, 
$t_1$ and $t_3$ parts are connected after omitting $t_2$ part 
to compare with the isothermal data at $T_{\rm L}$ (line). 
The average is from $6\times 10^3$ samples. 
 }
\label{fig:Pcycle}
\end{figure}

In Fig.~\ref{fig:Pcycle}, we show three typical behaviors observed 
in positive $T$-cycle simulations. In the main frame, we omit 
$t_2$ part of data and connect $t_1$ and $t_3$ parts to compare with 
the isothermal data drawn by line. 
For small $\textrm{NRG}$, $\chi_{\rm Pcycle}$ is 
remarkably below $\chi_{\rm iso}(T_{\rm L})$ 
in the third stage, as illustrated in the main frame of 
Fig.~\ref{fig:Pcycle}(a). This means that equilibration at $T_{\rm L}$ 
is sharply accelerated in the second stage because renormalized 
couplings at $T_{\rm L}$ and $T_{\rm H}$ are strongly correlated 
and equilibration is accelerated at $T_{\rm H}$. 
This trend (acceleration of equilibration) 
arises until $\textrm{NRG}\approx 8$. Fig.~\ref{fig:Pcycle}(b) 
shows that $\chi_{\rm Pcycle}$ begins to have a strong curvature 
in the third stage as a sign of rejuvenation, while renormalized couplings 
at $T_{\rm L}$ and $T_{\rm H}$ are still highly correlated. 
(As shown in Table~\ref{tab:Cvalues}, $C= 0.92$ for $\textrm{NRG}=9$.) 
However, $\chi_{\rm Pcycle}$ is still below $\chi_{\rm iso}(T_{\rm L})$ 
at later times of the third stage. Finally, Fig.~\ref{fig:Pcycle}(c) 
is the result for $\textrm{NRG}=13$. Renormalized couplings are 
very decorrelated now ($C= 0.10$). We see strong rejuvenation 
in the third stage, as is found in experiments. 
We have checked that rejuvenation is {\it perfect} 
in the sense that $\chi_{\rm Pcycle}$ in the first stage 
and that in the third stage completely overlap. 
We have also found that $\chi_{\rm Pcycle}$ in the second stage 
and $\chi_{\rm iso}(T_{\rm H})$ collapse into a single curve 
for any ${\rm NRG}$s, 
meaning that aging at lower temperature is not helpful in 
equilibration at higher temperature. 

\begin{figure}[ht]
\includegraphics[angle=0,width=\columnwidth]{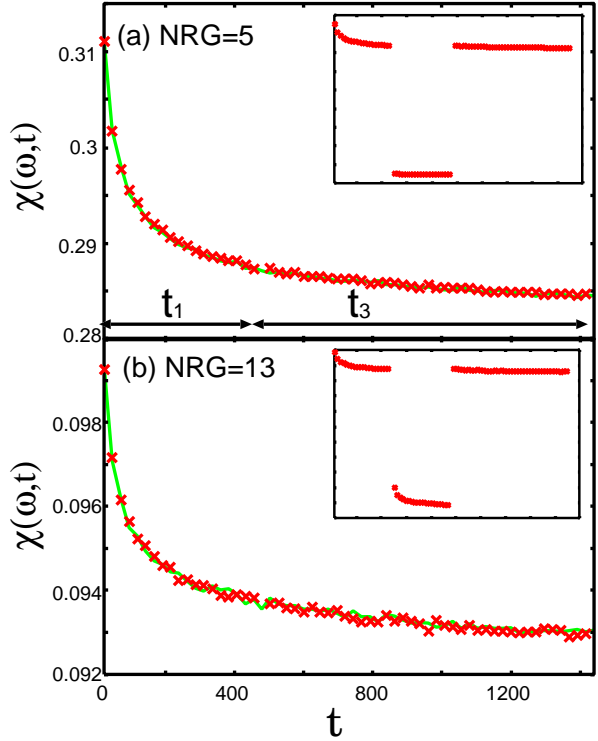} 
\caption{$\chi(\omega,t)$ with a negative $T$-cycle 
for $\textrm{NRG}=5$ and 13. Temperature is temporarily 
decreased from $T_{\rm H}=0.7$ to $T_{\rm L}=0.65$ during the period $t_2$. 
In the main frame, $t_1$ and $t_3$ parts are connected after omitting 
$t_2$ part to compare with the isothermal data at $T_{\rm H}$ (line). 
The average is from $6\times 10^3$ samples.}
\label{fig:Ncycle}
\end{figure}

For negative $T$-cycle case, we only show 
two extreme cases in Fig.~\ref{fig:Ncycle} , i.e., a highly correlated case and a highly decorrelated one. Now a surprising fact is that 
{\it perfect} memory is observed in the both cases. 
Especially, in the decorrelated case (${\rm NRG}=13$), memory appears 
in the third stage though rejuvenation is observed for 
positive $T$-cycle (Fig.~\ref{fig:Pcycle}(c)). 
This result is very contrast with~\cite{SasakiMartin03} which has shown 
that negative $T$-cycle leads to strong (but not perfect) rejuvenation 
when $\textrm{NRG}=\textrm{NRG}'$. These findings tell us that 
temperature dependence of speed of equilibration 
is crucial for memory effect. 
Lastly, we have compared $\chi_{\rm Ncycle}$ in the second stage 
with $\chi_{\rm Pcycle}$ in the third stage, and 
found that they perfectly overlap for any ${\rm NRG}$s. 
Since their difference is 
whether previous aging at $T_{\rm L}$ exists or not, 
this result suggests that equilibration at higher temperature 
makes previous aging at lower temperature insignificant. 

\paragraph*{Discussion and conclusions ---}
In this work, we have found that positive $T$-cycle
and negative one cause quite different effect, 
i.e., rejuvenation in the former and memory effect in the latter, 
if we take temperature dependence of speed of equilibration into account. 
In fact, the both lead to strong rejuvenation 
when renormalized couplings are decorrelated and 
$\textrm{NRG}=\textrm{NRG}'$~\cite{SasakiMartin03}. These results are 
interpreted as follows. Concerning negative $T$-cycle, 
what happens in the second stage is reconstruction of 
spin polarizations at shorter length scales. 
Since the structure at larger length scales 
(polarizations of spins with smaller generations in our model) 
created in the first stage is not destroyed at the time, 
we see memory in the following third stage. On the other hand, 
for positive $T$-cycle, the length scale reconstructed in the second stage is 
larger than the equilibrated length scale in the first stage. 
As a result, the order created in the first stage is completely destroyed, 
and we see rejuvenation. This scenario is very similar 
to the picture of~\cite{BouchaudDupuis01} where separation 
of the relevant length scale at each temperature plays a crucial 
role in memory and rejuvenation. However, the only difference is 
that temperature chaos exists for sure in MK spin glasses, 
while their picture relies on not temperature chaos but 
reweighing of Boltzmann factor by temperature variations 
as the cause of rejuvenation. 

Lastly, we comment on acceleration of equilibration observed 
in positive $T$-cycle for small $\textrm{NRG}$s. In experiments, 
this behavior is observed in many glassy systems like 
polymer glasses~\cite{BellonCiliberto99} 
and orientational glasses~\cite{AlberichDoussineau97}. 
Furthermore, the same behavior is also observed 
in spin glasses if the equilibrated length scale during the second stage 
is much smaller than $\ell(T,T')$. 
(See Fig.~14 of~\cite{JonssonYoshino03}.) These results may suggest that 
temperature chaos is absent, or the 
equilibrated length scale in experimental time scale 
is much smaller than $\ell(T,T')$ for these systems. 
\section{Acknowledgments}
M. S. was partially supported 
by the Japan Society for the Promotion of Science 
for Japanese Junior Scientists. 
The present simulations have been performed on SGI 2800/384 at the 
Supercomputer Center, Institute for Solid State Physics, 
the University of Tokyo. This work was also supported 
in part by the European Community under contract 
HPRN-CT-2002-000307 (Dyglagemem).
M. S. acknowledges support from the MENRT while he was in France. The 
LPTMS is an Unit\'e de Recherche de 
l'Universit\'e Paris~XI associ\'ee au CNRS.

\bibliographystyle{prsty}
\bibliography{references}

\end{document}